\documentclass[a4paper,fleqn,usenatbib,useAMS]{mnras}

\usepackage{graphicx}	
\usepackage{amsmath}	
\usepackage{amssymb}	
\usepackage{multicol}        
\usepackage{bm}		
\usepackage{pdflscape}	

\usepackage[T1]{fontenc}
\usepackage{ae,aecompl}

\usepackage{newtxtext,newtxmath}

\title[Magnetospherically-trapped dust and WD\,1145+017]{Magnetospherically-trapped dust and a possible model for the 
unusual transits at WD\,1145+017}

\author[J. Farihi, T. von Hippel,  J. E. Pringle]{
J. Farihi$^{1}$,
T. von Hippel$^{2,3}$, 
J. E. Pringle$^{3}$\thanks{E-mail: jep@ast.cam.ac.uk}
\\
$^{1}$Department of Physics and Astronomy, University College London, London WC1E 6BT, UK\\
$^{2}$Center for Space and Atmospheric Research, Embry-Riddle Aeronautical University, Daytona Beach FL 32114, USA\\
$^{3}$Institute of Astronomy, University of Cambridge, Cambridge CB3 0HA, UK}

\begin{document}


\maketitle

\begin{abstract}

The rapidly evolving dust and gas extinction observed towards WD\,1145+017 has opened a real-time window onto the mechanisms 
for destruction-accretion of planetary bodies onto white dwarf stars, and has served to underline the importance of considering the 
dynamics of dust particles around such objects.  Here it is argued that the interaction between (charged) dust grains and the stellar 
magnetic field is an important ingredient in understanding the physical distribution of infrared emitting particles in the vicinity of such 
white dwarfs.  These ideas are used to suggest a possible model for WD\,1145+017 in which the unusual transit shapes are caused 
by opaque clouds of dust trapped in the stellar magnetosphere.  The model can account for the observed transit periodicities if the 
stellar rotation is near 4.5\,h, as the clouds of trapped dust are then located near or within the co-rotation radius. The model requires 
the surface magnetic field to be at least around some tens of kG.  In contrast to the eccentric orbits expected for large 
planetesimals undergoing tidal disintegration, the orbits of magnetospherically-trapped dust clouds are essentially circular, 
consistent with the observations.  

\end{abstract}


\section{Introduction}

Evolved planetary systems orbiting white dwarfs provide unique and complementary information to conventional planetary system
studies.  Transit and radial velocity searches yield the frequency, minimum masses, and sizes for closely orbiting planets, where
small and likely rocky planets can sometimes be identified with confidence \citep{gil16}, but generally these data are insufficient
\citep{rog15}.  Despite the large datasets of planetary system frequencies, architectures, planet sizes and even densities, the only
empirical knowledge on planet compositions relates to their atmospheres, and only for those systems that are amenable to transit
or direct spectroscopy \citep{sin16,chi17}.  

The compositions of the most intriguing worlds, the small and rocky planets where knowledge of their surfaces is key to their 
habitability, remain firmly out reach via conventional observation.  Fortunately, white dwarfs accreting planetary debris are common 
\citep{zuc10,koe14}, as evidenced by the observed correlation between atmospheric heavy elements and infrared emission from 
closely-orbiting dust, usually taken to be in the form of disks \citep{von07,far09}.  These systems distill the infalling planetary debris 
via atmospheric pollution, and provide powerful insight into the mass and bulk chemistry of the parent bodies \citep{kle10,gan12}.  
For the handful of systems with detailed measurements, nearly all are consistent with distinctly terrestrial-like, differentiated parent 
bodies \citep{jur13,jur14}, a few objects show evidence for water or hydrated minerals \citep{far13,rad15,far16}, and one apparently 
ice-rich object \citep{xu17}.

This general picture is now rather compelling with more than a decade of corroborating, multi-wavelength observations \citep{gan06,
jur07}, including the detection of deep and irregular transits at WD\,1145+017\footnote{A modest quality spectrum of WD\,1145+017 
taken prior to 1990 appears to show Ca\,{\sc ii} K \citep{ber92}.  If correct, this system has been actively polluted for well over a quarter 
century.}  \citep{van15,gan16}.  Theoretical work has been carried out for disk evolution and accretion in these systems \citep{raf11,
met12}, and for post-main sequence, dynamical instabilities that may perturb planetesimals onto star-grazing orbits \citep{ver13,fre14}.  
But while simple models can tidally fragment bodies interior to the Roche limit \citep{deb12,ver14}, the resulting highly eccentric rings 
require Myr timescales or longer to shrink and circularize via stellar radiation \citep{ver15}, as the bulk of the debris mass for dust 
production will be contained in cm-size or larger particles \citep{wya08}.  One scenario likely involves mutual collisions -- especially 
near periastron -- and gas drag from sublimated material \citep{far12,bro17}.

The detection of transits towards WD\,1145+017 is consistent with the basic picture outlined above, as this star is dusty and enriched 
with numerous heavy elements \citep{van15}.  Quasi-periodic behavior is observed in both {\em K2} and ground-based transit light 
curves, mostly centered around 4.5\,h but with substantial variations (and uncertainty), and varying interpretations \citep{rap16,gar17,
cro17}.  It is noteworthy that individual bodies are not inferred to be the cause of the extinguished light, but rather sizable clouds of 
debris -- several times the size of the Earth \citep{gan16} -- that are presumed to be associated with orbiting planetary bodies or 
fragments.  Significant columns of metal-rich gas are also seen in absorption, and which vary on minute- to month-long timescales 
similar to the overall behavior of the light curve \citep{xu16,red17}.

A distinct challenge to the canonical model of tidal disruption is the apparent orbits of the transiting debris clouds.  The Keplerian 
orbit for a 4.5\,h period around a typical white dwarf is just inside 1.2\,$R_{\odot}$, but any significant eccentricity would result in 
catastrophic fragmentation for 1000\,km bodies \citep{bro17}.  For example, a rubble pile of this size and $\uprho \approx 3-4$\,g\,cm$
^{-3}$ should totally disrupt within 1\,yr unless $e\lesssim0.1$ \citep{ver17}, but the circularization of such an object is non-trivial by 
tidal forces \citep{ver16}.  

Motivated by these considerations and others, this paper investigates the possibility that such obscuring clouds are made of small 
dust grains, trapped in the white dwarf magnetosphere.  Section 2 outlines the physical processes required for dust grains to be 
strongly influenced by the stellar magnetic field.  It is shown that, in general, the smaller, infrared emitting grains are likely to be 
strongly affected by stellar magnetism for surface fields of around a few tens of kG.  In Section 3 these ideas are applied to 
WD\,1145+017, where it is proposed that the 4.5\,h photometric periodicities be co-identified with the stellar rotation period.  
An estimate is then made for the surface field strength required to ensure that small dust grains can be trapped close to the 
co-rotation radius.  Section 4 presents the discussion and conclusions.

\section{Dust Trapping Model}

Consider the influence of the white dwarf magnetic field on dust in its vicinity, where a typical relevant dust grain has size $a=1
\,\upmu$m, and hence a radius of $r_{\rm g}=a/2=0.5\,\upmu$m.  The initial assumption is that the grain is spherical.  This is 
likely a poor approximation and the effect of this assumption is considered later.  Taking the grain density to be $\uprho_{\rm g}=3
$\,g\,cm$^{-3}$, the mass of the dust grain is 
\begin{equation}
\label{grainmass}
m_{\rm g} = 1.6 \times 10^{-12}  \left( \frac{a}{1\,\upmu{\rm m}} \right)^3 \left( \frac{\uprho_{\rm g}}{3\,{\rm g\,cm}^{-3} } \right) \, {\rm g}.
\end{equation}

\subsection{Grain charge}

To quantify the interaction between dust grains and the stellar magnetic field, the charge on a typical dust grain needs to be
estimated, written here as $Ze$, where $-e$ is the charge of an electron.  The dust grains will likely be ionized by the impact of 
ultraviolet photons causing the expulsion of electrons \citep{hor96}.  The actual charge is therefore caused by a balance between 
the rate of impact of suitable photons and the rate at which electrons recombine with the dust grain.  It is assumed that the rate 
of impact of sufficiently energetic photons dominates, and that the grain reaches its maximum charge $Z_{\rm max}$ for which 
the electric potential of the grain surface prevents further expulsion of electrons by ultraviolet photons.

The model assumes the grain material has a work function of $\approx5$\,eV (see e.g.\ \citealt{hor96}).  Thus one can estimate 
that typical ultraviolet photons from the white dwarf with energies $\approx$10\,eV are able to expel electrons, provided 
that the net charge on the grain is such that the surface potential $\phi_{\rm g}$ is less than around 5\,eV.  These numbers are 
approximate, but will suffice for an initial estimate.  Again assuming the grain to be spherical, the surface potential is given 
by
\begin{equation}
\phi_{\rm g} = \frac{Z e}{ r_{\rm g}},
\end{equation}
where $Z$ is the charge on the grain in units of electron charge.

The spare energy given to the dislodged electron cannot exceed $e\phi_{\rm g}$, the energy needed to escape the surface positive 
charge, and thus the maximum value of $Z$ comes from setting $e \phi_{\rm g}=5$\,eV.  The charge on the grain is then
\begin{equation}
Z_{\rm max} = 1700 \left( \frac{a}{1\,\upmu{\rm m}} \right) \left( \frac{e \phi_{\rm g}}{5\,{\rm eV}} \right).
\end{equation}
It is noteworthy that this is only an estimate, and likely to be somewhat uncertain. For this reason the dependence on $e \phi_{\rm g}$, 
or equivalently on $Z_{\rm max}$, is carried through the analysis.

\subsection{Gyro-radius of the grains}

For the magnetic field to have an influence on the dust dynamics, to a first approximation the gyro-radius $R_{\rm G}$ of the dust grain 
must be comparable to, or smaller than, the radius at which the dust finds itself (cf.\ \citealt{ost13}).  The gyro-frequency is 
$\Omega_{\rm G} = ZeB/m_{\rm g} c$, and  thus the gyro-radius is given by
\begin{equation}
R_{\rm G} = \frac{m_{\rm g} c u_\perp}{Z e B},
\end{equation}
where $u_\perp$ is the particle velocity perpendicular to the field line.

\subsection{Field strength}

The model adopts a canonical surface magnetic field for the white dwarf of $B_*=10^3$\,G (cf.\ \citealt{azn04}), and a canonical 
white dwarf radius and mass of $R_*=10^9$\,cm and $M_*=0.6\,M_\odot$.  Assuming a dipolar field, at a distance of $r=1\,R_\odot$
the magnetic field strength would be
\begin{equation}
\label{Bfield}
B = 3.0 \times 10^{-3} \left( \frac{B_*}{1\,{\rm kG}} \right) \left( \frac{R_*}{10^9\,{\rm cm}} \right)^3 \left( \frac{r}{R_\odot} \right)^{-3} \, {\rm G}.
\end{equation}

\subsection{Velocity}

In the worst case scenario, the velocity of a dust grain will be the escape velocity
\begin{equation}
\label{escvel}
u = \left( \frac{2 GM}{r} \right)^{1/2} = 4.8 \times10^7 \left( \frac{M_*}{0.6\,M_\odot} \right)^{1/2} \left( \frac{r}{R_\odot} \right)^{-1/2} \, {\rm cm\,s}^{-1},
\end{equation}
with $u_\perp = f u$, and where the factor $f\le1$ tracks the assumptions about $u_\perp$.

\subsection{Magnetic dust radius}
\label{fdiscussion}

Putting the above together, the gyro-radius is now given by
\begin{equation}
\begin{split}
R_{\rm G} = 9.0 \times 10^{13} \left( \frac{\uprho_{\rm g}}{3\,{\rm g\,cm}^{-3}} \right) \left( \frac{a}{1\,\upmu{\rm m}} \right)^2 \left( \frac{r}{R_\odot} \right)^{5/2} f \left( \frac{M_*}{0.6\,M_\odot} \right)^{1/2} \\
\left( \frac{B_*}{1\,{\rm kG}} \right)^{-1} \left( \frac{R_*}{10^9\,{\rm cm}} \right)^{-3} \left( \frac{e \phi_{\rm g}}{5\,{\rm eV}} \right)^{-1} \, {\rm cm}.
\end{split}
\end{equation}
More useful is the ratio 
\begin{equation}
\label{dustradiusratio}
\begin{split}
\frac{R_{\rm G}}{r} = 1.3  \times 10^3 \left( \frac{\uprho_{\rm g}}{3\,{\rm g\,cm}^{-3}} \right) \left( \frac{a}{1\,\upmu{\rm m}} \right)^2 \left( \frac{r}{R_\odot} \right)^{3/2} f \left( \frac{M_*}{0.6\,M_\odot} \right)^{1/2} \\
\left( \frac{B_*}{1\,{\rm kG}} \right)^{-1} \left( \frac{R_*}{10^9\,{\rm cm}} \right)^{-3} \left( \frac{e \phi_{\rm g}}{5\,{\rm eV}} \right)^{-1}.
\end{split}
\end{equation}
Because $R_{\rm G}/r \propto a^2$, at a given radius, smaller grains are affected more strongly than larger ones.  

It is noteworthy that in general $f = u / u_{\rm esc} < 1$.  For example, grains on circular orbits have velocities $u_{\rm circ } = 
(1/\sqrt{2}) u_{\rm esc}$.  And in general grain trajectories will not be perpendicular to the magnetic field lines.  Thus typically 
one might expect that $f \simeq 1/2$.

For the trajectory of a grain to be substantially affected by the stellar magnetic field, the model requires and sets $R_{\rm G}/r\le1$.
A radius $r_{\rm magdust}$ can now be defined, within which the dust is strongly influenced by the field.  This yields
\begin{equation}
\begin{split}
\frac{r_{\rm magdust}}{R_\odot} =  8.4 \times 10^{-3} \left( \frac{\uprho_{\rm g}}{3\,{\rm g\,cm}^{-3}} \right)^{-2/3} \left( \frac{a}{1\,\upmu{\rm m}} \right)^{-4/3}  f^{-2/3} \\
\left( \frac{M_*}{0.6\,M_\odot} \right)^{-1/3} \left( \frac{B_*}{1\,{\rm kG}} \right)^{2/3} \left( \frac{R_*}{10^9\,{\rm cm}} \right)^{2} \left( \frac{e \phi_{\rm g}}{5\,{\rm eV}} \right)^{2/3}.
\end{split}
\end{equation}
From these estimates, it is clear that the stellar magnetic field is likely to be important on the relevant scales.  Grains with sizes 
in the range $a \sim 0.1-1\,\upmu$m have been inferred to dominate the observed {\em Spitzer} fluxes of dusty white dwarfs 
\citep{rea09,jur09}.  For such grains, with $f\simeq 0.5$ and surface magnetic fields of a few tens of kG \citep{azn04}, the stellar 
magnetosphere readily extends out to $r \approx 0.2\,R_\odot$, and for sub-micron grains to radii beyond $\approx1 \, R_\odot$.  These 
distance scales correspond to the inner regions of flat disk models \citep{far16b}, and the Roche limit \citep{ver14}, respectively, and 
within which dust production is likely to occur. 

However, there are two reasons why this is likely to be an underestimate.

\begin{enumerate}

\item 

{\em Grain shape.}  The above assumes that grains are spherical, and hence for a given mass this makes the grain as small as 
possible.  Yet, the observed polarization of starlight passing through dust demands that grains are non-spherical \citep{dra09}.  Thus, 
for non-spherical dust grains of a given mass, the effective radius $r_{\rm g}$ is underestimated.  This in turn implies an underestimate 
of the grain size relevant for ionizing photons, hence underestimating possible grain charge, because for a fixed surface potential, the 
larger the grain, the more charge it can hold.

Therefore, from consideration of the actual structure of grains, it is expected that $r_{\rm g}$ could be a factor of a few to several larger, or 
$B_*$ could be similarly smaller than implied by the above calculation.

\item

{\em Velocity -- the value of f.}  Taking the factor $f=1$ implies that the dust grain velocity relative to the magnetic field is the escape 
velocity ($u=u_{\rm esc}$), and that the dust grain is moving perpendicular to the magnetic field ($u=u_\perp$). As discussed above 
in Section~\ref{fdiscussion}, it is expected that $f < 1$ in general, and a more typical value is $f \simeq 0.5$.  This is the relevant value 
if the velocity were Keplerian, and if the dust particle were moving at 45$^\circ$ to the magnetic field.  All else being equal, the size of 
dust grains scales as $a\propto f^{-1/2}$. 

But likely the largest effect occurs if the white dwarf is rotating.  Thus for grains in a Keplerian orbit at radius $r = R_\Omega$, one
expects $f \ll1$. In other words, grains are most likely to be trapped by the field at radii close to where their intrinsic velocities are 
comparable to the local velocity of the magnetic field. Thus there is a strong tendency to trap co-rotating grains.

\end{enumerate}

Therefore stellar magnetic fields are likely to play an important role in the dynamics of dust accretion in many of these systems.

\section{Application to Obscurations at WD\,1145+017}

These ideas are now applied to obscuration events such as the observed light curve of WD\,1145+017, based on the simple 
idea that the dipping behavior is caused by clouds of dust.  To date, these individual clouds of dust have been assumed to originate 
from (potentially drifting) asteroid fragments in orbit around the star \citep{rap16}, where the $P\simeq4.5$\,h periodic behavior within 
the light curve has been interpreted as an orbital period for the asteroid fragments \citep{van15}.  A circular orbit with that period is 
located at a radius of $R_\Omega$ where 
\begin{equation}
\label{corotation}
R_\Omega = ( GM_*)^{1/3} \left( \frac {P}{2\uppi} \right)^{2/3} = \ 8.1 \times 10^{10} \left( \frac {M_*}{0.6\,M_\odot} \right)^{1/3} \left( \frac{P}{4.5\,{\rm h}} \right)^{2/3} \, {\rm cm}.
\end{equation}
From Equation \ref{dustradiusratio}, the smaller dust grains are those that are more likely to be affected by the stellar magnetic field.  
\citet{cro17} indicate that dust grains with sizes either $a > 0.30 \, \upmu$m or $a < 0.12 \, \upmu$m would fit their multicolor photometric 
data.  Based on this, a canonical dust grain size of $a = 0.1 \, \upmu$m is adopted here.

In order to obtain periodic obscuration signals, the model identifies the light curve period $P$ as the rotation period of the white dwarf,
and suggests that the dust clouds are trapped in the stellar magnetosphere. Ideally the dust would be trapped at or slightly interior to the 
co-rotation radius $R_\Omega$.  This is defined as the radius at which a circular Keplerian orbit has the same period as the stellar rotation 
\citep{pri72}.  An analogous model for periodic photometric variability has been proposed for rapidly rotating, weak-lined T\,Tauri stars in 
Upper Sco \citep{sta17}. 

The magnetospheric radius $R_{\rm B}$ is defined as the radius at which accreting material becomes attached to the magnetic field lines.  
The significance of the co-rotation radius is that if $R_{\rm B} \lesssim R_\Omega$, then once accreting material becomes attached to a 
{\em radial} field line it can still be accreted onto the star, because gravity exceeds centrifugal force.  However, if $R_\Omega \lesssim 
R_{\rm B}$, then the accretion flow can be disrupted by the centrifugal force once matter is attached to the field lines. 

These ideas were introduced originally in the context of gaseous disk accretion in X-ray binaries. In that case the definition of the 
magnetospheric radius is relatively straightforward, though somewhat assumption-dependent \citep{pri72,dav73,gho77}. However, 
for dust accretion such as considered here, the physics is more complicated.

For the proposed model for WD\,1145+017 to operate, $R_{\rm G} \approx k R_\Omega$ is required for grains with sizes in the region of 
interest, i.e.\ $a\simeq0.1\,\upmu$m.  The parameter $k$ is introduced, and expected to be of order unity, in order to keep track of the 
approximate nature of where relative to $R_\Omega$ the dust might be trapped.  Setting the left hand side of Equation \ref{dustradiusratio} 
to unity, and substituting $r= k R_\Omega$ from Equation \ref{corotation} into the right hand side of Equation \ref{dustradiusratio}, it follows that 
in order for the dust to be influenced by the magnetic field at the co-rotation radius, the required stellar magnetic field has magnitude
\begin{equation}
\begin{split}
B_* = 16 \left( \frac{\uprho_{\rm g}}{3\,{\rm g\,cm}^{-3}} \right) \left( \frac{a}{0.1\,\upmu{\rm m}} \right)^2  f \left( \frac{M_*}{0.6\,M_\odot} \right)  \left( \frac{R_*}{10^9\,{\rm cm}} \right)^{-3} \\
\left( \frac{e \phi_{\rm g}}{5\,{\rm eV}} \right)^{-1} k^{3/2} \left( \frac {P}{4.5\,{\rm h}} \right) \, {\rm kG}.
\end{split}
\end{equation}
Thus, for stellar surface fields of $B_*\approx10$\,kG and greater, one would nominally expect grains with sizes $a\lesssim 0.1\,\upmu$m 
to be affected at, or around, the co-rotation radius.

However, whilst it has been shown that a field of a few tens of kG can play a role in controlling the trajectories of individual dust particles, 
a further consideration is the field strength necessary to control the total mass of dust required to provide the observed obscuration towards 
WD\,1145+017.

As a starting point, \citet{cro17} suggest that the obscuring clouds are of size $R_{\rm o} \sim 9.5 \, R_\oplus \approx 6.1 \times 10^9$\,cm. 
That work focuses on grain sizes of order $a \sim 0.1 \, \upmu$m, and typical drops in flux during an event as being $\approx10$\,per cent. 
Using these numbers, the total number of grains in a typical cloud is
\begin{equation}
N \sim 1.5 \times 10^{29} \left( \frac{a}{0.1 \, \upmu{\rm m}} \right)^{-2}.
\end{equation}
Using Equation (\ref{grainmass}) for the mass of a grain, the mass of the cloud can now be estimated, and hence its mean density is
\begin{equation}
\uprho_{\rm cl} \approx 2.5 \times 10^{-16} \left( \frac{a}{0.1 \, \upmu{\rm m}} \right) {\rm g\,cm^{-3}}.
\end{equation}
In order that the field, $B$, is sufficiently strong to control such clouds at a radius $ r_{\rm cl} = k R_\Omega$, the requirement at that radius
is
\begin{equation}
B^2/4 \uppi \approx \uprho_{\rm cl} \, u_{\rm cl}^2,
\end{equation}
where $u_{\rm cl}$ is the velocity of the cloud, and $u_{\rm cl} \approx 2 \uppi r_{\rm cl}/P$ is the co-rotating velocity at that radius. Thus the
condition becomes
\begin{equation}
B(k R_\Omega) \approx 1.8 k \left(   \frac{a}{0.1 \, \upmu{\rm m}} \right)^{1/2} \left( \frac{M_*}{0.6\,M_\odot} \right)^{1/3} \left( \frac{P}{4.5\, {\rm h}} \right)^{-1/3}  {\rm G}.
\end{equation}
This would correspond to a field at the stellar surface given by $B_* = B(k R_\Omega) (k R_\Omega / R_*)^3$, that is
\begin{equation}
\begin{split}
B_* \approx 940 \, k^4 \left(   \frac{a}{0.1 \, \upmu{\rm m}} \right)^{1/2} \left( \frac{M}{0.6 \, M_\odot} \right)^{4/3} \left( \frac{P}{4.5 \, {\rm h}} \right)^{5/3}  \\ \left( \frac{R_*}{ 10^9 \, {\rm cm}} \right)^{-3} {\rm kG}.
\end{split}
\end{equation}
The conclusion is that for a suitable set of system parameters, a plausible model for the periodic photometric behavior of WD\,1145+017 
could be based on the concept of dust clouds trapped near the magnetospheric co-rotation radius.  As noted above, analogous behaviour 
is seen elsewhere in stellar systems \citep{sta17}.

The major problem with this speculative model is with the strength of required surface field.  A field of some tens of kG is able to influence 
the trajectories of individual dust particles out to the co-rotation radius. However, such a field is not able to control dust clouds necessary 
to provide the observed obscuration out to that radius.  However, if the geometry is such that the necessary clouds are able to accumulate 
at radii slightly smaller that the co-rotation radius, such that for example $k \approx 0.3 - 0.5$, then a surface field of some tens of kG would 
again be sufficient.

\section{Discussion and Conclusions}

This work has shown that the interaction between dust and stellar magnetospheres is likely to be an important physical effect in 
contributing to the understanding of the dynamics of circumstellar debris around white dwarf stars. Such considerations are already 
known to be important in the understanding of accretion of gaseous material onto many other types of stellar object including:

\begin{enumerate}

\item accretion onto magnetic neutron stars as models for X-ray pulsars \citep{lew06};

\item accretion onto magnetic white dwarfs as models for AM\,Her stars (polars) and DQ\,Her stars (intermediate polars) 
\citep{war95};

\item accretion onto T Tauri stars \citep{joh07}.

\end{enumerate}

In all these cases the interaction between the accretion flow and the stellar field at the magnetospheric radius is poorly understood, 
but the interactions are known to be highly unstable and result in variable accretion flux from the stellar surface \citep{pri72,dav73,
aro76a,aro76b,gho77,aro80}. In the cases of white dwarfs \citep{kui82} and of neutron stars \citep{mor84} it has been suggested 
that clumps form in the flow at the magnetosphere, and that these clumps arrive as distinct entities at the stellar surface.

In the case discussed here, the interaction at the magnetosphere is even more complicated because the dust flow is likely not 
hydrodynamic and is more akin to a marginally collisional plasma. Estimates of the mean free paths of the relevant dust grains can 
be obtained by considering the optical depths of the clouds, and the tendency of the small grains to form clouds may be related to 
the shortened mean free paths in density enhancements. Nevertheless, predicting the nature (sizes and masses) of the dust clouds 
that are likely to form is beyond the scope of this paper.

In summary, it has been demonstrated that a stellar magnetic field of a few tens of kG can strongly influence the dynamics of dust 
particles within the tidal truncation zone. Moreover, it is important to note that these magnetic interactions are most important for the 
dust particles with sizes $a\lesssim1\,\upmu$m, which are those likely to contribute most to the observed infrared emission. The 
larger and more massive dust particles -- carrying the bulk of the debris mass -- are essentially unimpeded. 

\subsection{WD1145+017}

These general ideas have been used to construct a tentative model to explain the periodic photometric behaviour of the polluted 
white dwarf WD\,1145+017. For this model to work, the following is required:

\begin{enumerate}

\item The stellar rotation period must be approximately equal to the period of the observed dipping events, that is $P\approx4.5$\,h.
There are currently more than one dozen isolated white dwarfs with rotation periods of less than 5\,h, as measured via periodic light
curve modulations or asteroseismology (e.g.\ \citealt{her17}).  Intriguingly, a significant fraction of these are magnetic \citep{kaw15}, 
but this may be a selection bias as starspots readily allow the detection of rotation period, which is otherwise challenging for the 
bulk of white dwarfs.

\item The line of sight to the white dwarf needs to lie in or near the (rotational) equatorial plane of the white dwarf. This is because 
the clouds of dust held up by centrifugal force at the magnetosphere are likely to accumulate close to the equatorial plane. Similarly, 
those dust particles trapped in the magnetosphere, but closer to the spin axis, do not feel the full effects of centrifugal force and are 
able to accrete more easily. Note that because of the nature of the scattering process which is thought to bring the planetesimal debris 
into the sphere of influence of the central white dwarf, there is no expectation that the spin axis of the white dwarf and the spin axis of 
any putative dust disk be aligned (e.g.\ \citealt{fre14}).

\item In order to significantly affect the dust dynamics, the surface field of the white dwarf need only be modest, preferably greater 
than around $B_*  \simeq 10 - 50$ kG.  Fields of such a magnitude should be detectable with spectropolarimetry (e.g.\ \citealt{lan12,
lan16}), although such a detection would not be straightforward given the amount of variable absorbing material around the star, and 
given that for high-resolution spectra (e.g.\ \citealt{xu16}) integration times can be comparable to the stellar rotation period.

\end{enumerate}

\section*{Acknowledgements}

The authors thank D. Wickramasinghe for insights into the measurements of white dwarf magnetic fields, as well as N. Achilleos 
and J. J. Hermes for input.  J. Farihi acknowledges funding from the STFC via an Ernest Rutherford Fellowship, and the Institute 
of Astronomy for visiting support during the preparation of the manuscript.  The authors thank the referee for helpful comments.

\bsp    
\label{lastpage}
\end{document}